\documentstyle[preprint,prl,aps,epsf]{revtex}
\begin{document}
\preprint{\vbox{\hbox {May 1998} \hbox{IFP-758-UNC} } }
%\twocolumn[\hsize\textwidth\columnwidth\hsize\csname@twocolumnfalse\endcsname

%\draft
\title{\bf Bileptons from Muon Collider Backward Scattering.}
\author{\bf Paul H. Frampton and Xiaohu Guan}
\address{Department of Physics and Astronomy,}
\address{University of North Carolina, Chapel Hill, NC  27599-3255}
\maketitle
\begin{abstract}
There are serious discussions for building a muon collider with
$\mu^{+}\mu^{-}$ collisions at c.o.m. energies up to $4$ TeV.
We point out that the bileptonic gauge bosons predicted in some
extensions of the Standard Model would be readily discernable from
the backward scattering cross-section, for bilepton masses up to
a substantial fraction of the c.o.m. energy.
\end{abstract}
%\vskip0pc]
\newpage
In some extensions of the Standard Model, there occur
additional gauge bosons with lepton number $L = \pm 2$ which are called bileptons. 
Some reviews of their properties are given in \cite{a,b}.
The detailed couplings of bileptons to fermions
depend on the model. 
Generically they appear in $SU(2)$ doublets $(Y^{++},Y^{+})$ and the
charge conjugates, and couple to two leptons with a gauge coupling $g$
of order the electron charge $e$ but with the precise value depending on the
model. In some models, bileptons couple also to quarks but this will 
not be relevant here.

One means of discovering bileptons which has been discussed\cite{c,d}
is to look for the direct channel resonance in an electron linear
collider in the $e^{-}e^{-}$ mode. For a muon collider the $\mu^{-}\mu^{-}$
mode could obviously be used similarly with no
special advantage.

The principal advantage of a muon collider over an electron collider
is that because of
the much lower synchrotron radiation a "circular" machine can be used and
c.o.m. energies up to several TeV can be foreseen. 
With this possibility, our purpose here is to point
out what is quite simple: that for $\sqrt{s} >> M_Y$,
and a coupling $g \geq e$ the backward scattering produced
by bileptons can compete with the forward Coulomb peak.
We shall illustrate this with the explicit
lowest-order Born calculations of the cross-section and corresponding
Figures for c.o.m. energies in the range $500$ GeV to $4$ TeV and
bilepton masses $M_Y$ from $300$ GeV to $2$ TeV.

The current lower limits on the bilepton masses are
$M(Y^{+}) > 230$ GeV from polarized muon decay\cite{e} and
$M(Y^{++}) > 360$ GeV from muonium-antimuonium conversion\cite{f}.

The Feynman rule for coupling a doubly-charged bilepton
to two like-sign muons arises from the interaction:
\begin{equation}
\frac{g}{\sqrt{2}} Y_{\mu}^{++} \mu^{T} C \gamma_{\mu}
\left( \frac{1 - \gamma_5}{2} \right) \mu 
+
\frac{g}{\sqrt{2}} Y_{\mu}^{--} \bar{\mu} \gamma_{\mu}
\left( \frac{1 - \gamma_5}{2} \right) C \mu^{T}
\end{equation}

In the $SU(15)$ model the value of the coupling is $g = 1.19e$ \cite{g}
while in the $331-$ model the coupling is somewhat larger: $g =
2.085e (= e/sin\theta)$\cite{h}. We shall consider both possibilities.

The differential cross-section for $\mu^{+}\mu^{-} \rightarrow
\mu^{+}\mu^{-}$ can be written\cite{g} 

\newpage

\begin{eqnarray}
\frac{d\sigma (\mu^{+}\mu^{-} \rightarrow \mu^{+}\mu^{-})}{d cos\theta}
& = &\frac{\pi \alpha^2}{2s} \left[ \left[|G_{LR}(t)|^2 +
|G_{RL}(t)|^2\right] +\left(\frac{t}{s}\right)^2\left[|G_{LR}(s)|^2 + 
|G_{RL}(s)|^2\right]\right. \nonumber \\ 
&   &+ \left. \left(\frac{u}{s}\right)^2\left[|G_{LL}(s) + G_{LL}(t)|^2
+ |G_{RR}(s) + G_{RR}(t)|^2\right]\right] \label{cs}
\end{eqnarray}
where
\begin{equation}
G_{AB} (w) = \frac{s}{w} + \frac{g_Ag_B}{e^2}\frac{s}{w - M_Z^2}
+ (\delta_{AB} - 1) \left( \frac{g}{\sqrt{2} e}\right)^2 \frac{s}{u - M_X^2}
\end{equation}
with $A,B = L,R$. For the Standard Model (SM) piece,
$g_L = -e cot 2\theta_W, g_R = e tan\theta_W$. The SM
cross section is obtained for $g = 0$, the $SU(15)$ for
$g = 1.19e$ and the $331$ model for $g = 2.085e$.

The formulae for $e^{+}e^{-} \rightarrow e^{+}e^{-}$ are 
identical to the above, so our focus
on $\mu^{+}\mu^{-} \rightarrow \mu^{+}\mu^{-}$ at the highest c.o.m.
energies is merely pragmatic.

In the above formulae, $s = (p_1 + p_2)^2, t = (p_1 - p_3)^2,
u = (p_1 - p_4)^2$ for $\mu^{+}(p_1) + \mu^{-}(p_2) \rightarrow
\mu^{+}(p_3) + \mu^{-}(p_4)$, and $\theta$ is the scattering angle between
initial and final $\mu^{+}$ in the c.o.m. frame.

The results are shown in the Figures. For the
331-model we have used $M_Y = 300, 500$ and $700$ GeV
(in the simplest minimal version, $M_Y \leq 800$ GeV\cite{i}
although with an enlarged Higgs sector involving
an octet of $SU(3)_L$ this bound can be relaxed).
Figs. 1 and 2 then show the differential cross sections
for $\sqrt{s} = 1$ and $4$ TeV, respectively.

For the $SU(15)$ model where $M_Y$ has no sharp upper
bound, in Figs. 3 and 4 we have used 
$\sqrt{s} = 1$ TeV and $\sqrt{s} = 4$ TeV, respectively.
For $\sqrt{s} = 1$ TeV we set $M_Y = 300, 500, 700$ GeV
and $1$ TeV, while for $\sqrt{s} = 4$ TeV we plot also $M_Y
= 1.5$ and $2$ TeV.

All the Figures also show the Standard Model differential
cross section.

For the relatively light bilepton masses $M_Y = 300 \sim 700$ GeV
we see from Figs. 1 and 2 firstly that the back-scattering effect is very large.
Also, we see that, as expected, the distinction between the low masses
may be easier at the low c.o.m. energy
$1$ TeV than at $4$ TeV. On the other hand, the
ratio of the differential cross section of the 331-model 
to the Standard Model becomes larger as $\sqrt{s}$ 
is increased.

Similar conclusions hold for the $SU(15)$ case: the
overall effect is down by a factor
$(g_{15}/g_{331})^4 \sim 0.106$ but is still readily distinguishable.
Here bilepton masses up to $2$ TeV (disallowed in the minimal
331-model) can be distinguished. 

Finally, we mention polarized $\mu^{+}\mu^{-}$ scattering.
From Eq.(\ref{cs}) we see that $(d\sigma/dcos\theta)$ for
LR (= RL) polarizations is proportional to that
for the unpolarized case in the backward direction for
which $u = 0$ and $t \simeq -s$, so there is
no special advantage of polarization for this discovery process.

This work was supported in part by the U.S. Department of Energy under 
Grant No. DE-FG05-85ER-40219.

\newpage

{\bf Figure Captions.}

1. Differential cross-section 
$d\sigma(\mu^{+}\mu^{-}\rightarrow\mu^{+}\mu^{-})/d cos\theta$
in 331-model for $\sqrt{s} = 1$ TeV and bilepton masses
$M_Y = 300, 500$, and $700$ GeV. 

2. Differential cross-section 
$d\sigma(\mu^{+}\mu^{-}\rightarrow\mu^{+}\mu^{-})/d cos\theta$
in 331-model for $\sqrt{s} = 4$ TeV and bilepton masses
$M_Y = 300, 500$, and $700$ GeV. 

3. Differential cross-section 
$d\sigma(\mu^{+}\mu^{-}\rightarrow\mu^{+}\mu^{-})/d cos\theta$
in $SU(15)$ model for $\sqrt{s} = 1$ TeV and bilepton masses
$M_Y = 300, 500, 700$ GeV and $1$ TeV. 

4. Differential cross-section 
$d\sigma(\mu^{+}\mu^{-}\rightarrow\mu^{+}\mu^{-})/d cos\theta$
in $SU(15)$ model for $\sqrt{s} = 4$ TeV and bilepton masses
$M_Y = 300, 500, 700$ GeV and $1, 1.5, 2$ TeV.

\newpage
\begin{figure}
\begin{center}

\epsfxsize=6.0in

\ \epsfbox{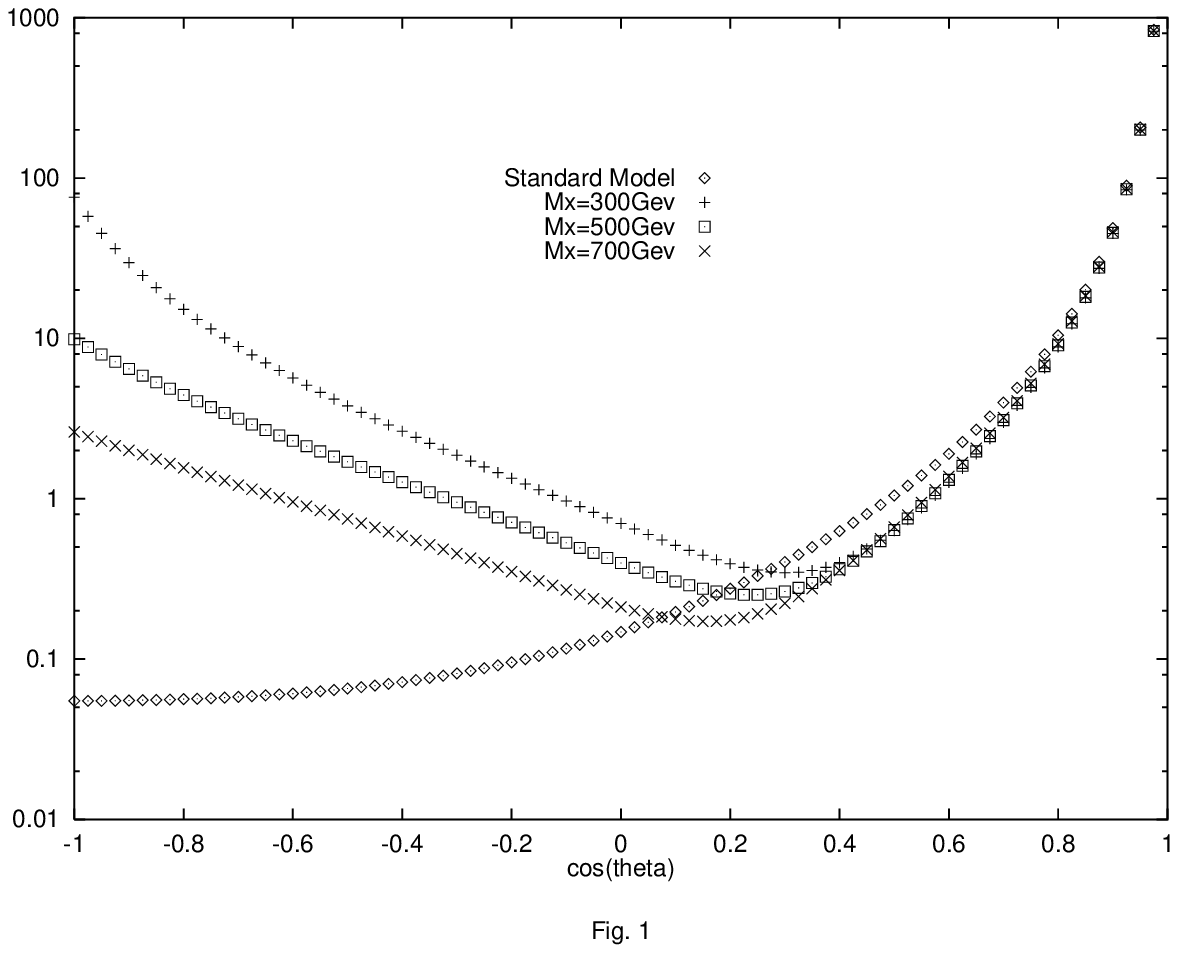}

\end{center}
\end{figure}

\newpage
\begin{figure}
\begin{center}

\epsfxsize=6.0in

\ \epsfbox{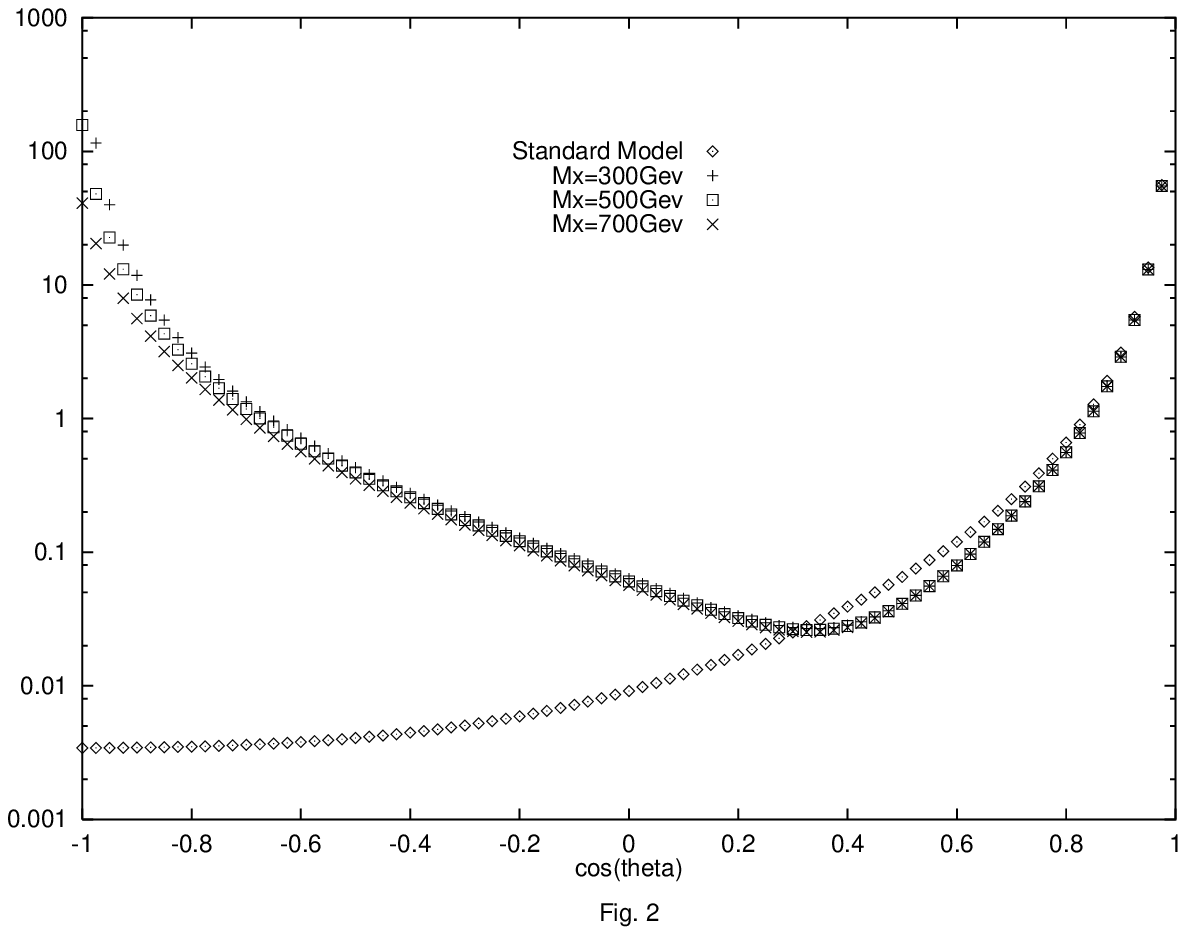}

\end{center}
\end{figure}

\newpage
\begin{figure}
\begin{center}

\epsfxsize=6.0in

\ \epsfbox{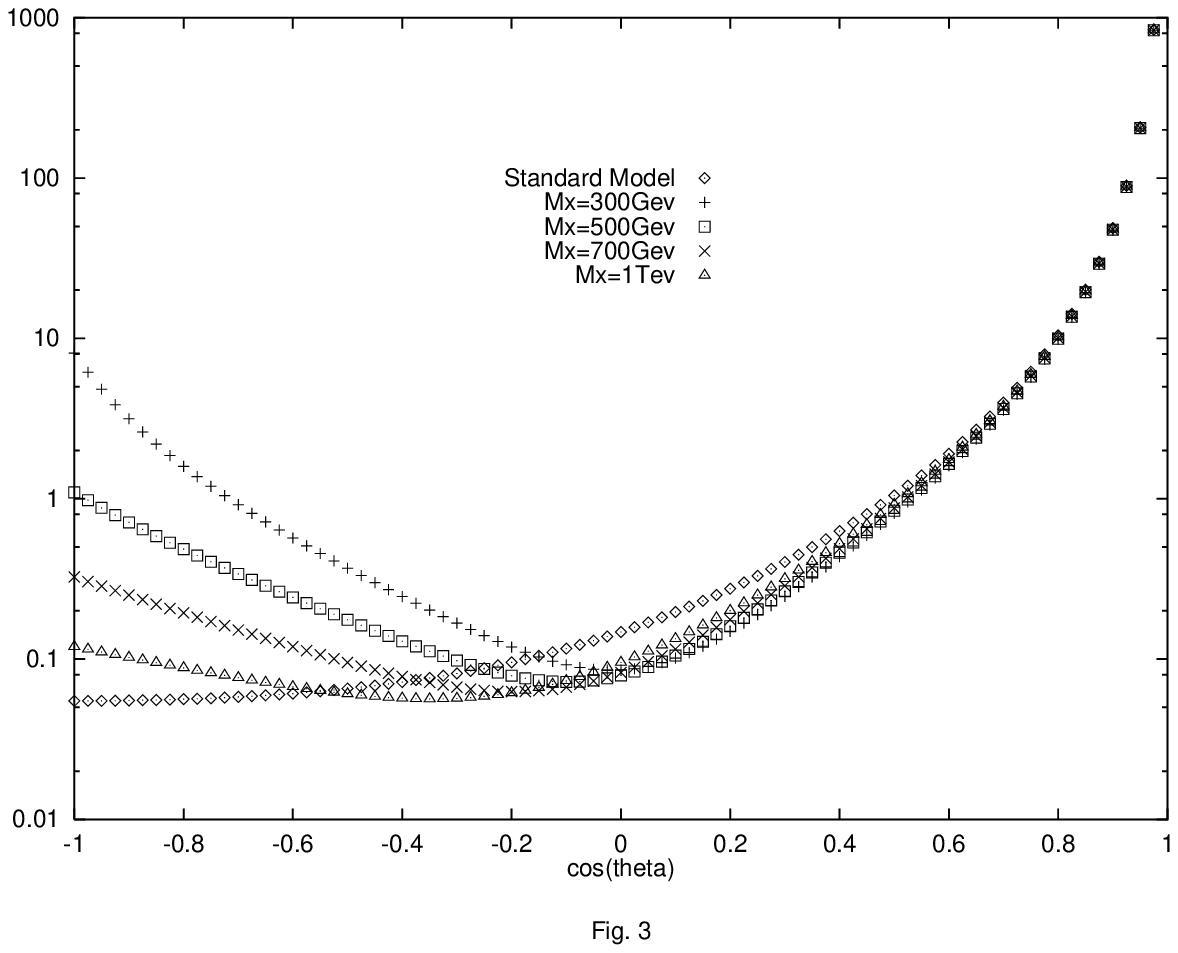}

\end{center}
\end{figure}

\newpage
\begin{figure}
\begin{center}

\epsfxsize=6.0in

\ \epsfbox{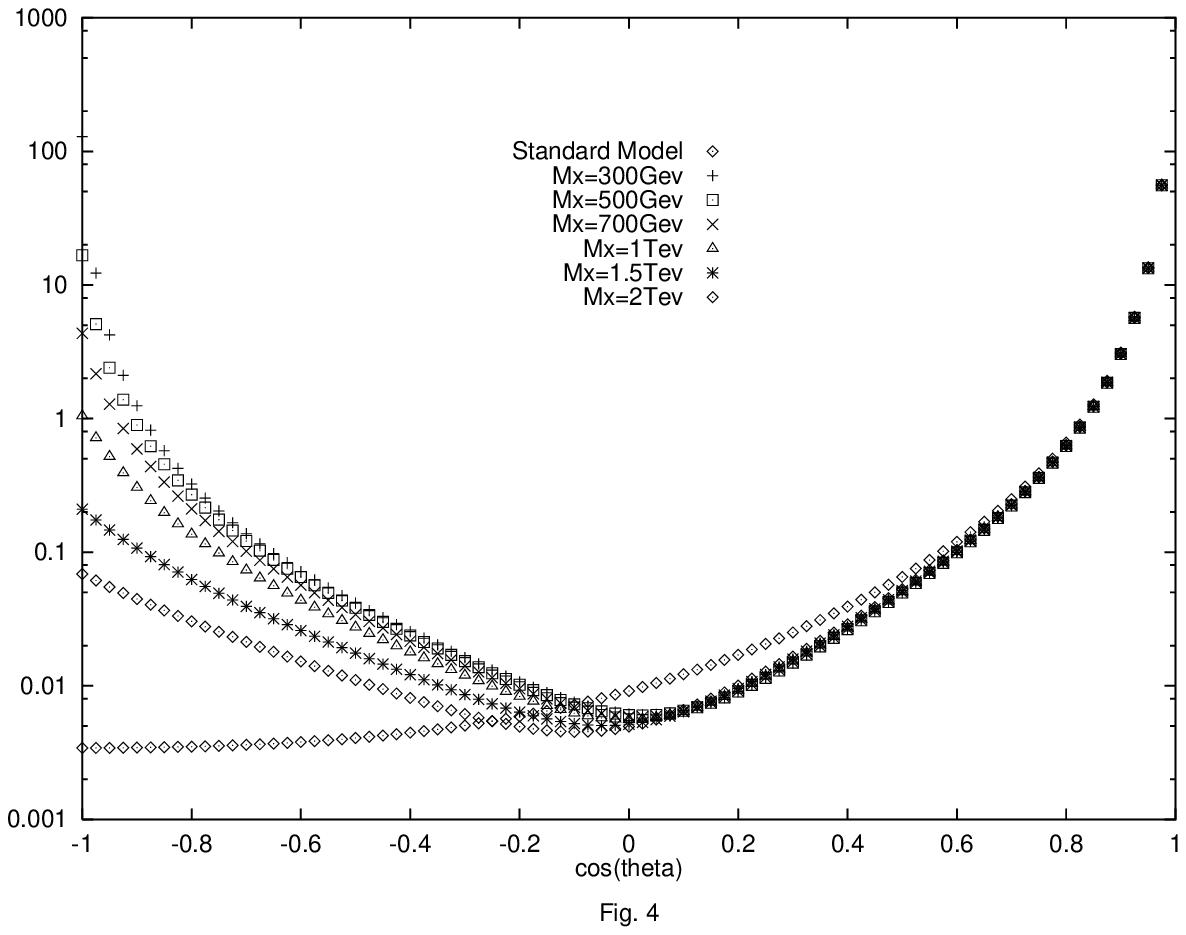}

\end{center}
\end{figure}

\end{document}